\documentclass[lettersize,journal]{IEEEtran}
\usepackage{amsmath,amsfonts}
\usepackage{algorithmic}
\usepackage{algorithm}
\usepackage{array}
\usepackage[caption=false,font=normalsize,labelfont=sf,textfont=sf]{subfig}
\usepackage{textcomp}
\usepackage{stfloats}
\usepackage{url}
\usepackage{verbatim}
\usepackage{graphicx}
\usepackage{cite}
\usepackage{doi}
\usepackage{color}
\begin{document}

\raggedbottom

\title{Generative AI-Driven Hierarchical Multi-Agent Framework for Zero-Touch Optical Networks}

\author{Yao Zhang, Yuchen Song, Shengnan Li, Yan Shi, Shikui Shen, Xiongyan Tang, Min Zhang, Danshi Wang
\thanks{This work was supported by National Natural Science Foundation of China (No. 6252200718, U24B20133, 62171053), Beijing Nova Program (No. 20230484331), and BUPT Excellent Ph.D. Students Foundation (No. CX20241033) (Corresponding author: Danshi Wang).}
}

\markboth{ACCEPTED BY IEEE COMMUNICATIONS MAGAZINE, 2025 }%
{Shell \MakeLowercase{\textit{et al.}}: A Sample Article Using IEEEtran.cls for IEEE Journals}


\maketitle

\begin{abstract}
The rapid development of Generative Artificial Intelligence (GenAI) has catalyzed a transformative technological revolution across all walks of life. As the backbone of wideband communication, optical networks are expecting high-level autonomous operation and zero-touch management to accommodate their expanding network scales and escalating transmission bandwidth. The integration of GenAI is deemed as the pivotal solution for realizing zero-touch optical networks. However, the lifecycle management of optical networks involves a multitude of tasks and necessitates seamless collaboration across multiple layers, which poses significant challenges to the existing single-agent GenAI systems. In this paper, we propose a GenAI-driven hierarchical multi-agent framework designed to streamline multi-task autonomous execution for zero-touch optical networks. We present the architecture, implementation, and applications of this framework. A field-deployed mesh network is utilized to demonstrate three typical scenarios throughout the lifecycle of optical network: quality of transmission estimation in the planning stage, dynamic channel adding/dropping in the operation stage, and system capacity increase in the upgrade stage. The case studies, illustrate the capabilities of multi-agent framework in multi-task allocation, coordination, execution, evaluation, and summarization. This work provides a promising approach for the future development of intelligent, efficient, and collaborative network management solutions, paving the way for more specialized and adaptive zero-touch optical networks.
\end{abstract}

\begin{IEEEkeywords}
Generative AI, Multi-Agent, Zero-Touch Optical Networks.
\end{IEEEkeywords}

\vspace{-4mm}

\section{Introduction}
\IEEEPARstart{G}{enerative} Artificial Intelligence (GenAI) refers to a type of advanced AI techniques capable of generating meaningful and contextually relevant content in multiple modalities, including text, image, audio, video, and multimodal combinations. Unlike traditional AI models that focus on classification, regression, or prediction tasks, GenAI systems equipped with reasoning capabilities can create novel outputs by learning from extensive and diverse data sources. GenAI leverages transformer-based large foundation models to perform various tasks, like content generation, reasoning, translation, summarization, and prediction. In recent years, GenAI has expanded its application scope through the integration of advanced technologies, such as large language models (LLMs), visual language models (VLMs), video diffusion models (VDMs), and multimodal fusion techniques. The rapid development and upgrades of GenAI have triggered a surge of research and applications across diverse domains \cite{ref1}, driven by their human-like reasoning and problem-solving abilities. This progress promoted industries to not only automate traditionally labor-intensive tasks, but also tackle some long-standing challenges \cite{ref2}, ushering in an era of technological integration and innovation.

Recently, GenAI has been integrated with specialized domain knowledge to explore its promising applications in zero-touch networks \cite{ref3}. Notably, several LLM-driven single-agent solutions have been developed to address specific tasks in optical network, including network planning \cite{ref4}, resource allocation \cite{ref5}, alarm and log analysis \cite{ref6, ref7}, failure handling \cite{ref8}, device configuration \cite{ref9, ref10}, and performance optimization \cite{ref11}. Basically, these LLM-based single-agent solutions have demonstrated commendable performance in addressing individual and well-defined sub-tasks through well-crafted prompt engineering or domain-specific fine-tuning techniques. However, when applied to full lifecycle management of optical networks, the existing single-agent solutions encounter significant challenges in simultaneously handling multiple intricate, cross-layer, and large-scale tasks, hindering the realization of fully autonomous operation and comprehensive zero-touch management.

Specifically, firstly, the one-size-fits-all characteristic of general-purpose GenAI models often prove inadequate for solving several highly specialized problems within the optical network domain. To address it, current techniques such as prompt engineering, model fine-tuning, retrieval augmented generation have been employed to customize the specialized models and proven effective in solving individual problems. However, the need for a framework that is both globally adaptable and domain-specialized remains a pressing challenge, as no single tuned model can seamlessly address all interrelated tasks automatically. Secondly, the lack of collaborative mechanism and comprehensive management fundamentally restricts existing single-agent systems from functioning as a cohesive and unified system. Instead of relying on a single agent for all tasks, the multi-agent frameworks that utilize a set of specialized models to intricate problems have been proposed as a paradigm shift for large-scale and complex systems with the goal of enhancing performance, scalability, and flexibility \cite{ref12}. Accordingly, it is foreseeable that multi-agent framework offers a promising solution for achieving full automation and comprehensive management of optical networks through multi-agent seamless collaboration and strict strategy alignment. Finally, task execution in optical networks is characterized by the diversity in complexity, modalities, and priority, necessitating varying model scales, domain-specific training data, hierarchical user permissions, and specialized functional capabilities. Multi-agent framework allows for the deployment of professional GenAI-based experts within a well-structured system, which is designed with clearly defined agent roles, distinct capabilities, hierarchical permissions, and tailored reward mechanisms.

In this paper, we proposed the integration of GenAI within a hierarchical multi-agent framework to drive the autonomous operations for zero-touch optical network. Multiple LLM-based AI Agents are specifically customized to execute task allocation, coordination, implementation, evaluation, and summarization. We utilized the GenAI models to facilitate seamless communications of workflows based on natural language, while maintaining high levels of task precision using the hierarchical multi-agent framework. Through experimental validation in field-deployed optical mesh network, we demonstrate the effectiveness of the approach with three typical cases covering the network planning, operation, and upgrade stages, offering promising results in zero-touch management. In addition, we discussed some possible capability extensions to enable the development of the multi-agent framework into a more intelligent, adaptable, and comprehensive system that can meet the needs of the next generation of optical networks.

\begin{figure*}[ht]
\centering
\includegraphics[width=13cm]{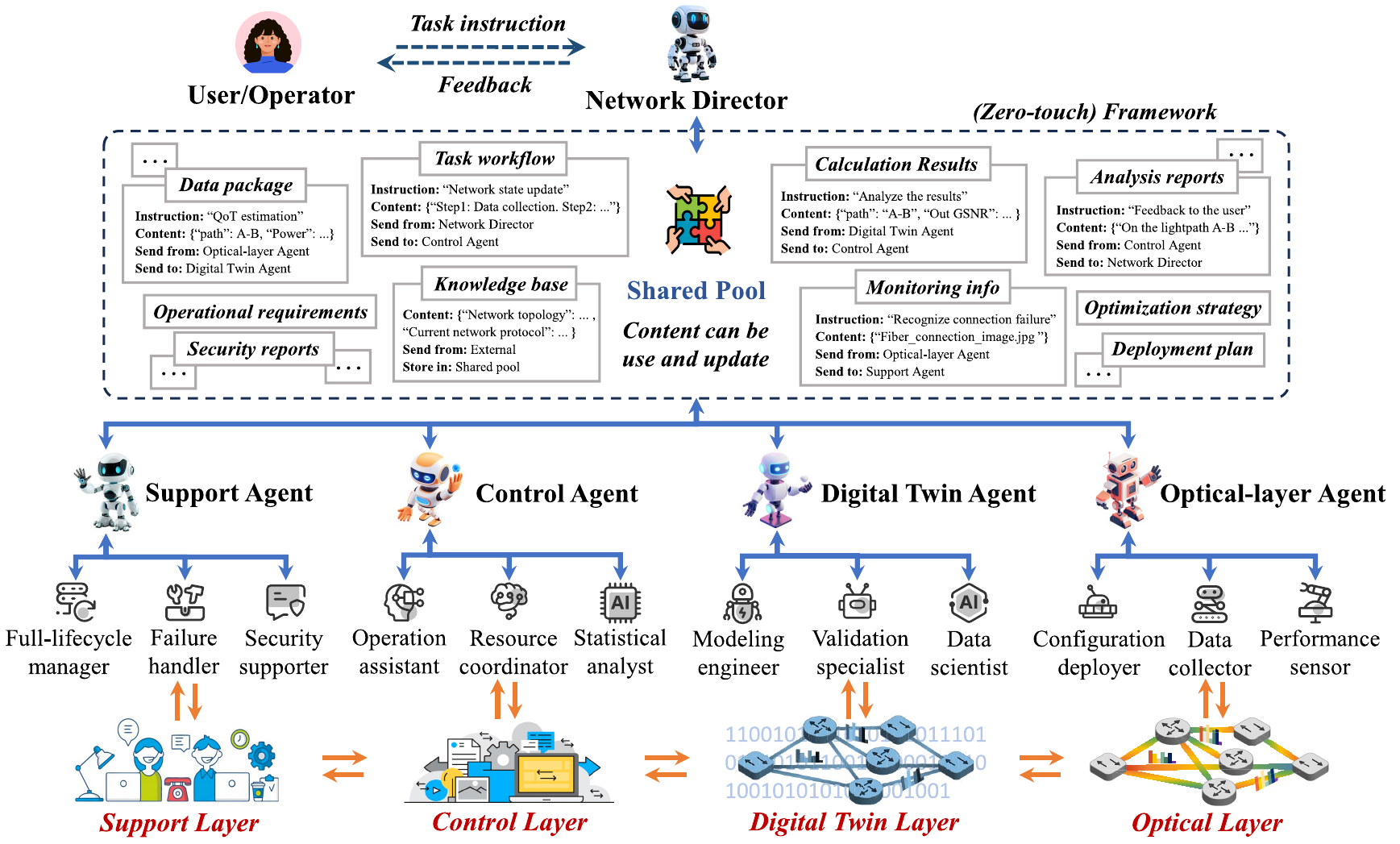}
\caption{GenAI-enabled hierarchical multi-agent framework, including the Network Director, Shared Pool, Division Agents and AI Experts cross four network layers: optical layer, digital twin layer, control layer and support layer.}
\label{fig_1}
\end{figure*}

\vspace{-3mm}

\section{GenAI-Driven Hierarchical Multi-Agent Framework}
\vspace{-1mm}

The proposed hierarchical multi-agent framework is constructed upon a multi-layer network structure, which is systematically subdivided into distinct hierarchical levels, as illustrated in Fig. \ref{fig_1}. The system is globally governed by a centralized agent at top level, while being coordinately managed by multiple specialized agents at different operational levels. Each agent is driven by GenAI and plays customized roles with interactive permissions to facilitate efficient task management, specialization, and coordination, ensuring optimal task execution, data transfer, and resource allocation. A pivotal component of this framework is the ``Shared Pool'', a middle interactive medium that serves as a repository for all the task-related contents. The Shared Pool is dynamically utilized and updated by the agents, providing a robust storage and caching mechanism that supports interaction between subdivisions and headquarter across layers. In this section, we mainly describe the hierarchical multi-agent framework and its workflow tailored for zero-touch optical networks.

\vspace{-4mm}

\subsection{Customization of Agents}
\vspace{-1mm}

The multi-agent framework needs function customization for each AI Agent to ensure the effective execution of tasks within their designated responsibilities, where each agent is composed of 1) an LLM backend for reasoning and decision-making, 2) a role-specific knowledge base for domain expertise, and 3) a tool interface for executing simulations, analysis, and network control actions. The AI Agent in the top level is the ``Network Director'', operating as the central brain for task allocation and coordination for the entire system, as shown in Fig. \ref{fig_1}. It is first customized with the basic knowledge of network operation to control the task execution process of the entire network automation. It serves as the primary contact point for users and has the highest permissions to access the Shared Pool, ensuring the security of the whole framework. 

In the middle level, we design four ``Division Agents'': Optical-layer Agent, Digital Twin (DT) Agent, Control Agent, and Support Agent, corresponding to four critical layers of optical network, including optical layer, DT layer, control layer, and support layer. Each Division Agent is tasked with managing specific domains or functional areas within their respective divisions. To optimize their performance, we meticulously customize each agent according to their unique work requirements, allocated resources, operational processes, and expected outputs. The Division Agents can access the Shared Pool to receive task workflow from the Network Director and obtain the data package, knowledge, and information required for their work. They directly guide ``AI Experts'' in the bottom level, assign sub-tasks, and monitor progress within their divisions, and they can review the outputs submitted from AI Experts before uploading them to the Shared Pool, to ensure accuracy and completeness.

\begin{figure*}[ht]
\centering
\includegraphics[width=12cm]{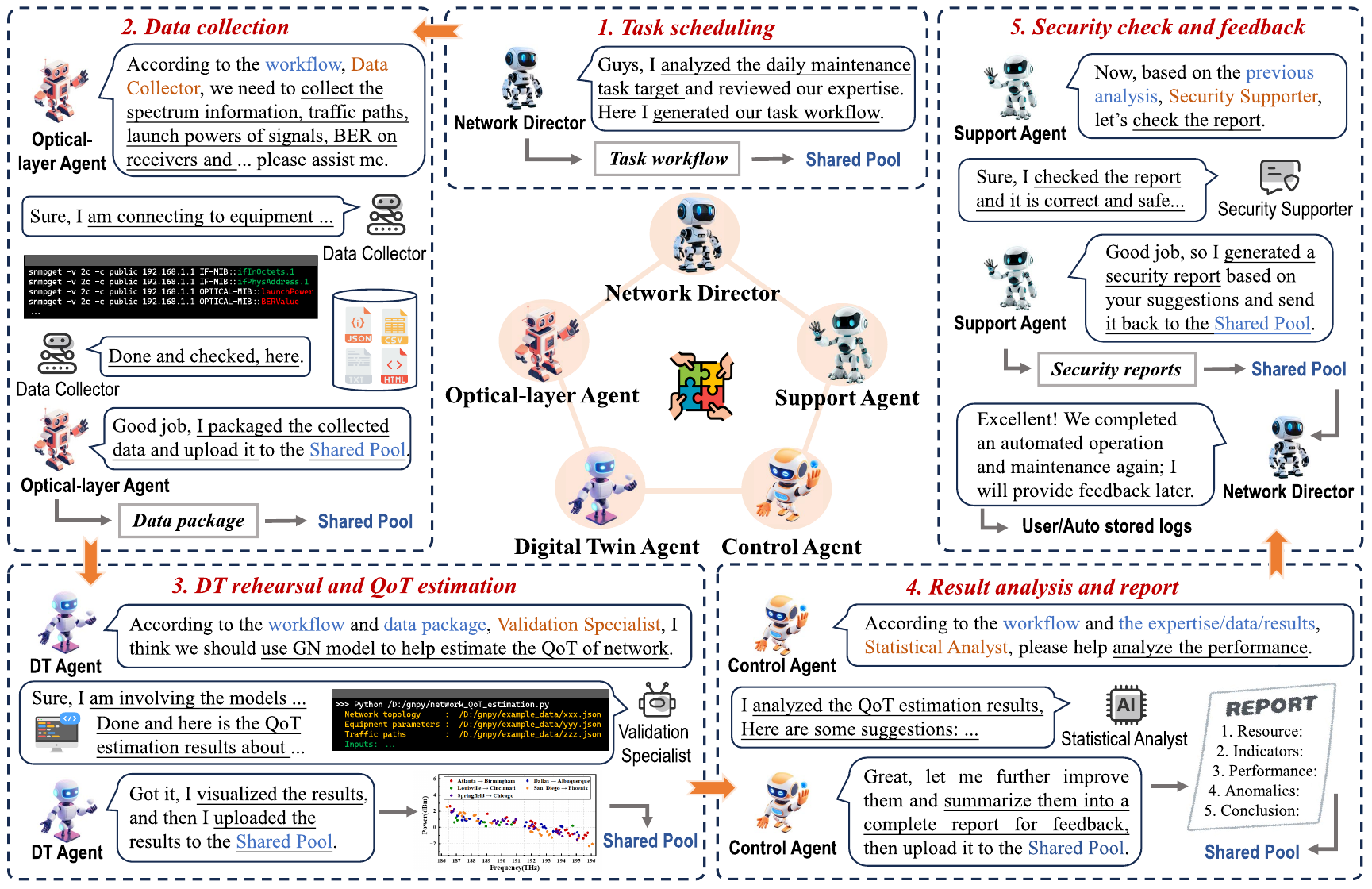}
\caption{The interaction and workflow among multi-agents for most daily maintenance tasks in the zero-touch optical network.}
\label{fig_2}
\end{figure*}

At the bottom level of the hierarchy, the agents designed as ``AI Experts'' focus on executing detailed and specific sub-tasks within their professional domains. They work under the direct supervision of Division Agents, following precise instructions and returning completed sub-task outputs. These AI Experts act as deployers, executors, workers, engineers, and supporters according to specific customizations to implement respective functions. The orchestration is built on custom Python modules with core agents can access and update the Shared Pool, while AI Experts, operating as lightweight models, directly interface with domain-specific tools. External tools such as enhanced GNPy for QoT estimation, K-shortest path and First-fit algorithms for routing and spectrum allocation, margin calculators, and application programming interfaces (APIs) for network management system (NMS)/software-defined network (SDN) controller are wrapped as callable Python functions to support task execution. For Division Agents, each division is controllable and scalable. We can continuously add new AI Experts, and the Division Agents just need to be fine-tuned to manage new experts, rather than learning detailed solutions.

Each layer is equipped with at least three AI Experts at the bottom level to execute the tasks required for optical network operation and maintenance. In the optical layer, the Configuration Deployer is designed to interface with the NMS for equipment configuration updates; the Data Collector aims to access NMS for performance data collection; the Performance Sensor is utilized for optical performance monitoring and link condition sensing. In the DT layer, the Modeling Engineer can create and maintain DT models to simulate the network’s behavior; the Validation Specialist aims to use these models for strategy rehearsal and algorithm validation; the Data Scientist is responsible for data management and secure utilization within the DT layer. In the control layer, the Operation Assistant supports daily network operations with suggestions; the Resource Coordinator manages resource allocation (including software and hardware resources); the Statistical Analyst provides comprehensive network status reports and actionable insights for improvement. Lastly, in the support layer, the Full-lifecycle Manager oversees network development planning; the Failure Handler resolves network alarms and failures events; the Security Supporter verifies the authenticity, integrity, and policy compliance of configuration instructions before approving their deployment to the physical layer. All of these AI Experts work together in their respective roles to ensure efficient and effective network operation.

\vspace{-4mm}

\subsection{Interaction among Agents}
\vspace{-1mm}

In the hierarchical framework, the interaction among multiple agents is essential and important. To enable this, we implemented the Shared Pool as a central repository that supports the memory and interaction by providing the agents with context-aware access to operational data and knowledge beyond the token limits of LLMs. It facilitates seamless communication and resource sharing among agents, including: task workflow scheduled by Network Director, data package collected from optical layer, knowledge base involved in basic network operation, calculation results outputted by DT layer, monitoring information extracted from the environment, analysis reports summarized by the Control Agent, and so on. All entries in the Shared Pool are stored in a consistent format, including instruction (sub-task target), shared content, sending and receiving objects, allowing efficient retrieval and clear permission management.

There are also cross-level interactions between Division Agents and AI Experts, reflected in the sending of instructions and the return of results. Considering the polysemous characteristics of natural language, a standardized interaction format is strictly defined between AI Experts and Division Agents for each transaction. Note that AI Experts can directly interact with their Division Agents, and work with the corresponding system, models, APIs, databases, textual materials, etc. This structure ensures that the task execution remains efficient, collaborative, and scalable while maintaining distinct role functions among the agents.

\vspace{-4mm}

\subsection{Workflow across Agents}
\vspace{-1mm}

To illustrate the workflow of this hierarchical multi-agent framework in network operations, the schematic diagram of multi-agent coordination for solving most daily maintenance tasks in the zero-touch optical network is displayed in Fig. \ref{fig_2}. First, the Network Director analyzes the task targets, leveraging its expertise to generate a comprehensive task workflow. This workflow specifies sub-task goals, responsible divisions, and execution sequences. It encompasses several steps: data collection, DT rehearsal, result analysis, and security check, which are then uploaded to the Shared Pool. Subsequently, the corresponding Division Agents proceed to execute their allocated sub-tasks sequentially.

The Optical-layer Agent starts by commanding the Data Collector to interface with the NMS and obtain real-time data. Next, the DT Agent analyzes and decides to use the Gaussian Noise (GN) model for quality of transmission (QoT) estimation. Then the Validation Specialist is instructed to invoke the DT for performance predictions. Following this, the Control Agent retrieves the data and results, and collaborates with Statistical Analyst to generate a response report. Based on this report, the Support Agent guides the Security Specialist in conducting network security analysis and final checks. Finally, the Network Director consolidates all results from the Shared Pool, and completes the task. This structured and collaborative process achieves the efficiency and adaptability of the hierarchical multi-agent framework for autonomous operations of zero-touch optical network.

\begin{figure*}[ht]
\centering
\includegraphics[width=13.5cm]{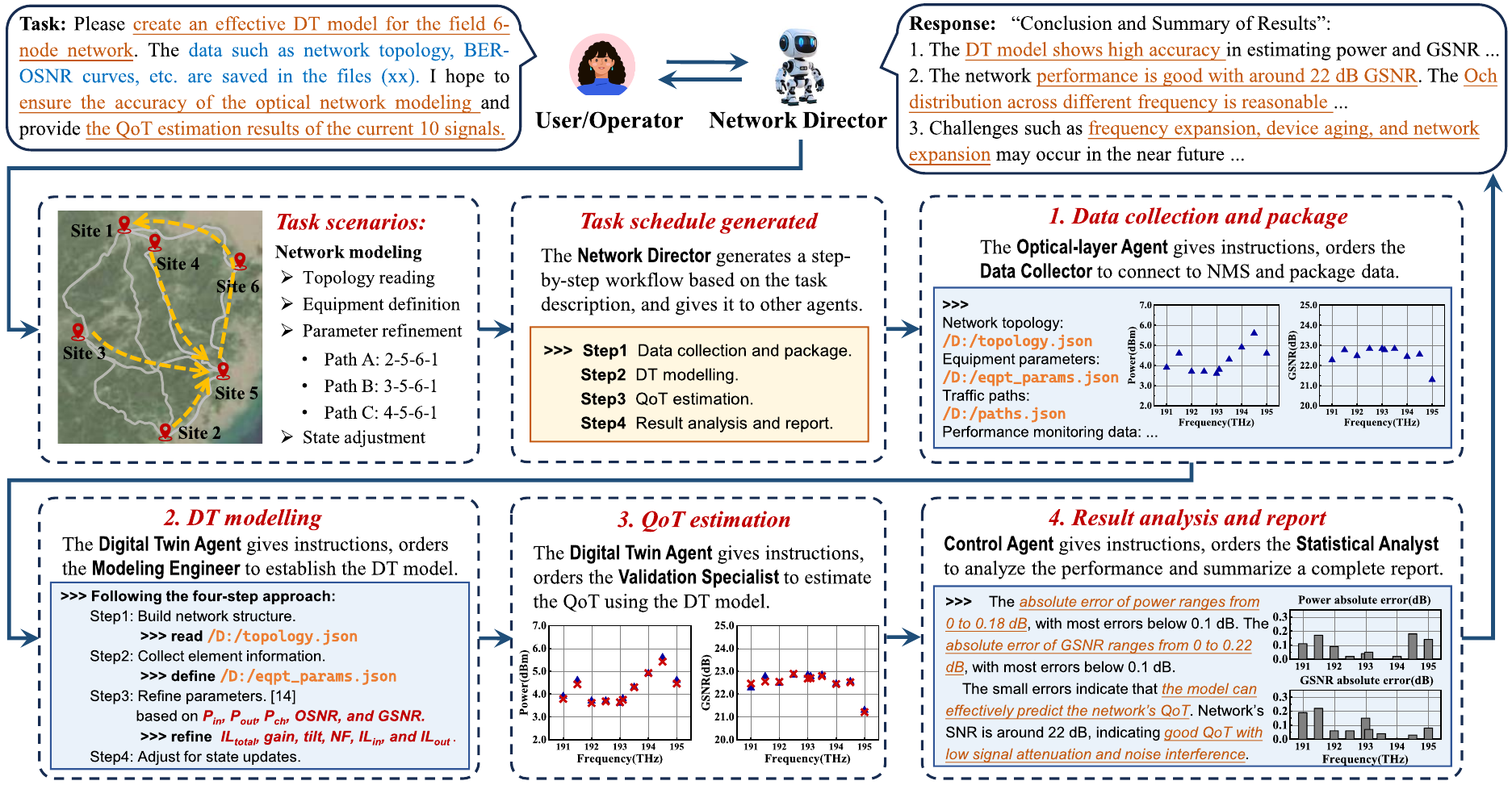}
\caption{Case 1: demonstration of a network planning scenario, using the multi-agent framework to assist in DT building and QoT estimation for service deployment.}
\label{fig_3}
\end{figure*}

\vspace{-2mm}

\section{Automatic Task Coordination in a Field-deployed Optical Network}

To evaluate the capability of the hierarchical multi-agent framework in automating task coordination within real-world scenarios, we conducted a series of field trials based on data collected from a field-deployed optical mesh network. The case study covered three typical scenarios throughout the lifecycle of optical network: 1) DT building and QoT estimation in the early planning stage, 2) dynamic channel adding/dropping in the regular operation stage, and 3) system capacity increase in the follow-up upgrade stage. We used the multi-agent framework to ensure network performance evaluation and effective implementation of planning, operating, upgrading works.

\vspace{-4mm}

\subsection{Setup}
\vspace{-1mm}

The field-deployed optical network consisted of 6 reconfigurable optical add-drop multiplexer (ROADM) sites deployed at 6 cities in China, as shown in the map of Fig. \ref{fig_3}. The length of standard single mode fibers (SSMFs) ranged from 47.0 km to 114.0 km, with Erbium doped fiber amplifier (EDFAs) deployed in each span for loss compensation. At the beginning, each transmitter-receiver (TRX) was equipped with the commercial 400Gb/s transponders for a total of 60 channels with 100-GHz spacing covering the whole C-band. The modulation format of signals was 16 quadrature amplitude modulation (QAM). In the following three cases, we set different channel allocations and deployed 8--10 channels in each case without adding other channels, making the channel adding/dropping-induced fluctuation more pronounced in some scenarios. 

In the verifications, the GPT-4o \cite{ref13} with the powerful abilities of intent comprehending and logic reasoning was utilized as the basic GenAI model for the multi-agent framework, which tended to answer in a more fluent and complete style. It could accept any combination of text, audio, and images as input and generate multimodal combined outputs. The techniques of chain-of-thought (CoT) prompt were adopted to customize multiple GPT-4o models designed for multiple agents corresponding to different roles. In our demonstrations, the agents were operated offline with collected data of the field-deployed network. Then, we assembled the multi-agent framework for the following case study.

\vspace{-4mm}

\subsection{Case 1: DT building and QoT estimation in the early planning stage}
\vspace{-1mm}

In the first case, we attempted to use the multi-agent framework to assist in DT building and QoT estimation for service deployment in a network planning scenario. The results of the proposed approach using real optical network data were displayed in Fig. \ref{fig_3} in the form of task flows. We deployed ten signals at different starting nodes on three paths with a frequency range of 191.0--195.0 THz. Although the beginning optical multiplex section (OMS) was different for these three sets of signals, all of them were multiplexed (without regeneration) through the same OMSs: Site 5--6--1 in the practical mesh network. Fig. \ref{fig_3} depicts the actual dataflow and the corresponding system responses under test conditions.

\begin{figure*}[ht]
\centering
\includegraphics[width=17.5cm]{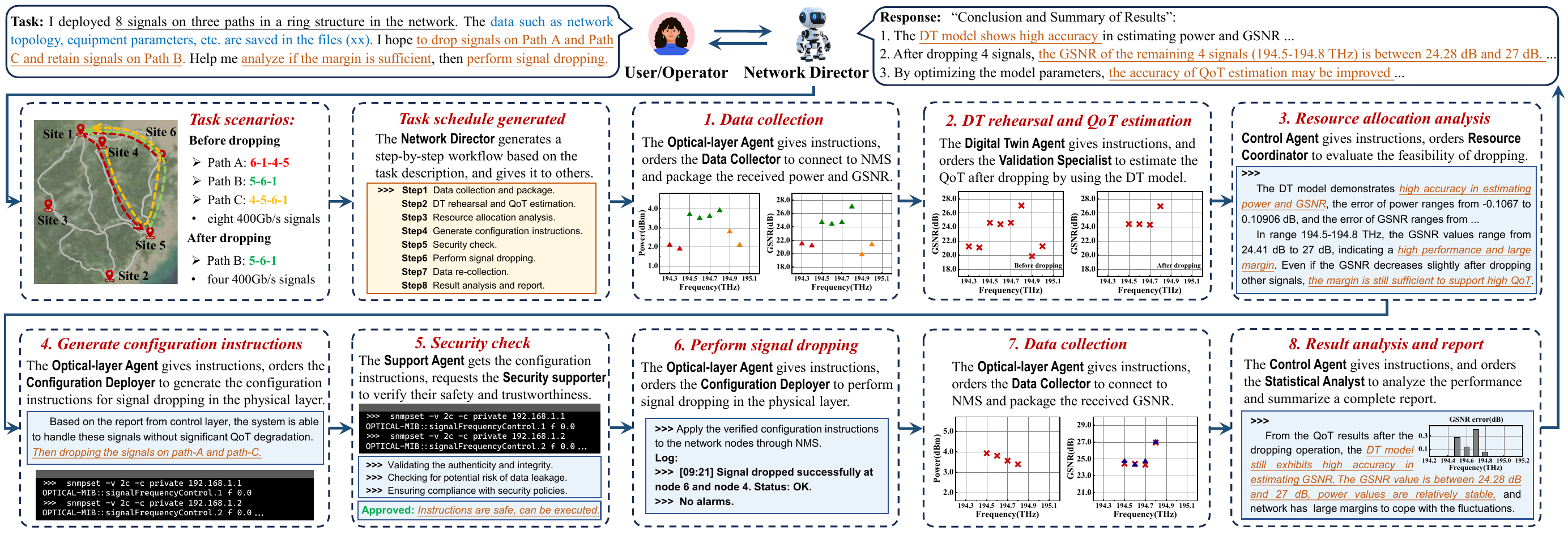}
\caption{Case 2: demonstration of a regular network operation scenario, using the multi-agent framework to assist in analyzing network performance and executing dynamic channel dropping.}
\label{fig_4}
\end{figure*}

The designated task target was ``\textit{to ensure the accuracy of the optical network modeling and provide the QoT estimation results of the current 10 signals}''. After receiving the task target, the Network Director first analyzed the task scenarios, and generated a step-by-step workflow. The task target was divided into four steps and executed sequentially: 1) Data collection and package: collecting the necessary data for modeling, along with the current network performance across three paths in the optical layer, and packaging them into the Shared Pool. 2) DT modeling: performing optical network modeling based on existing data. The Modeling Engineer in DT layer employed a four-step approach \cite{ref14} with a pre-set workflow to refine and adjust the parameters in sequence until an accurate DT model was established. 3) QoT estimation: utilizing the established DT model to estimate QoT within the DT layer. 4) Result analysis and report: analyzing the network performance and providing effective reports to the Network Director. The conclusion was that the current performance reflected a stable and healthy operation, with GSNR around 22 dB and margins acceptable; meanwhile, the deployment of the DT model demonstrated its effectiveness in accurately predicting QoT, with GSNR absolute errors remaining below 0.22 dB.

It should be noted that we visualized the task scenarios and execution results of each step for better understanding, and key contents in the analyzed text were underlined for highlight. In the actual execution process, these steps were performed automatically. Finally, the Network Director provided the operator with a response in natural language based on the information in the Shared Pool, thereby responding to the initial task target. This includes: 1) the accuracy and effectiveness evaluation of the established DT model; 2) the current performance evaluation of the practical optical network; and 3) prediction results and suggestions of the potential network operational risks in the future. 

\vspace{-4mm}

\subsection{Case 2: Dynamic channel adding/dropping in the regular operation stage}
\vspace{-1mm}

In the next case, we used the multi-agent framework to assist in regular network operation and maintenance, implementing a workflow for dynamic channel dropping, as shown in Fig. \ref{fig_4}. It was anticipated that the multi-agent framework could effectively handle such a complex task while ensuring network performance before and after the signal dropping. Eight signals were deployed in a ring route in the optical mesh network, with frequency range of 194.3--195.0 THz, and transmitted and received at four different nodes in the ring. We defined three paths, namely Path A, Path B, and Path C; then, we planned to drop the signals on Path A and Path C, and only retain the signals on Path B without offline.

\begin{figure*}[h]
\centering
\includegraphics[width=17.5cm]{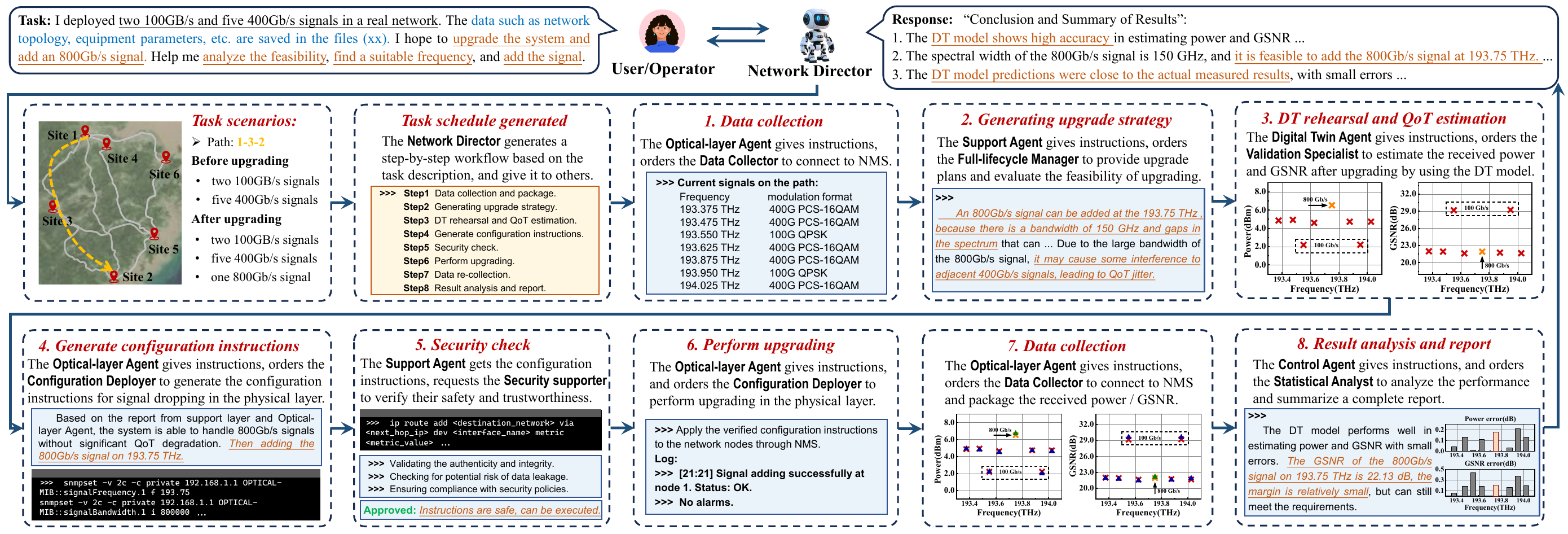}
\caption{Case 3: demonstration of a network upgrading scenario, using the multi-agent framework to assist in adding an 800Gb/s signal at an appropriate channel, while two 100Gb/s and five 400Gb/s signals have already existed on the same path.}
\label{fig_5}
\end{figure*}

The designated task target was ``\textit{to analyze whether the margin is sufficient for dropping signals on Path A and Path C and retaining signals on Path B, and if the conditions are met, to perform signal dropping}''. Then, the Network Director analyzed the task scenario and generated an eight-step workflow to execute: 1) Data collection and package: collecting the current performance on three paths in the optical layer and packaging them. 2) DT rehearsal and QoT estimation: utilizing DT tools to estimate QoT within the DT layer and predicting the network state after dropping four signals. 3) Resource allocation analysis: after analyzing QoT results, margin, spectrum, and fluctuations within the control layer, the Resource Coordinator providing the analysis that it was feasible for signal dropping. 4) Generate configuration instructions: based on the analysis results, Configure Deployer generating instructions for signal dropping. 5) Security check: checking authenticity and integrity of the instructions by Security Supporter, to ensure that can be applied. 6) Perform signal dropping:  executing signal dropping action through the Configure Deployer in the optical layer. 7) Data re-collection: collecting updated network performance. 8) Result analysis and report: analyzing the results and reporting that the network had enough margins to cope with the fluctuations caused by signal dropping, with GSNR ranging from 24.28 dB to 27.0 dB; meanwhile, the DT model still exhibited high accuracy in estimating GSNR, with absolute errors within 0.40 dB.

Similarly, the Network Director provided the operator with a response to the initial task target, including 1) the performance evaluation of optical networks before and after signal dropping; 2) error analysis between predicted QoT results and actual tested values; and 3) feasible optimization suggestions for performance fluctuations caused by the signal dropping.

\vspace{-3mm}

\subsection{Case 3: System capacity increase in the follow-up upgrade stage}

In the third case, we focused on capacity increase for network upgrading, specifically an 800Gb/s signal was expected to be placed at an appropriate channel, while two 100Gb/s and five 400Gb/s signals had already existed on the same path, as shown in Fig. \ref{fig_5}. In this upgrading task, utilizing multi-agent framework could provide outstanding assistance for coordinating tools to accurately simulate network behavior and predict the impact of a new signal on overall system performance, ensuring minimal interference with existing services. The designated task target was ``\textit{to help analyze the feasibility for upgrading the system, find an appropriate channel to add an 800Gb/s signal}''. Then, the Network Director analyzed target and generated an eight-step workflow to execute: 1) Data collection and package: collecting the current deployed information of two 100Gb/s and five 400Gb/s signals. 2) Generating upgrade strategy: creating an appropriate strategy by the Full-lifecycle Manager in the support layer, who provided reasons and instructed to add an 800Gb/s signal at 193.75 THz. 3) DT rehearsal and QoT estimation: rehearsing the upgraded network state within the DT layer. 4) Generate configuration instructions: Configure Deployer generating upgrade-related configuration commands. 5) Security check: Security Supporter verifying the upgrade instructions following security policies. 6) Perform upgrading: based on the reasons and predicted results, executing the upgrading action by the Configure Deployer. 7) Data re-collection: collecting updated network performance. 8) Result analysis and report: analyzing the results and reporting that through effective rehearsal, an 800Gb/s signal was successfully adding on 193.75 THz, with a received GSNR of 22.13 dB, ensuring minimal impact on other channels and meeting the initial requirements.

Finally, the Network Director provided the operator with a response to the initial task target, including feasible reasons before system upgrade and network performance after adding the 800Gb/s signal, as well as the evaluation of the DT model. From the operator's perspective, they only need to interact with the central brain (the Network Director) of the network, without having to juggle intricate task processes and tool usage, enabling them to directly obtain task execution results. Most importantly, the hierarchical multi-agent framework can provide very detailed and explainable responses for complex tasks step-by-step.

\vspace{-2mm}

\section{Discussions and Prospects}

\subsection{Efficiency and scalability}

The proposed hierarchical multi-agent framework was designed with modularity, scalability, and adaptability. In our tests across three representative cases (planning, operation, upgrade), task completion times remained below about 20 seconds, demonstrating low latency enabled by the hierarchical structure. It integrates data collection, and a DT of optical network, allowing continuously dynamic rehearsal and accurate performance estimation. As networks expand in scale or functionality, new divisions or specialized agents can be seamlessly integrated, while lightweight local intelligence deployed at the edge reduces coordination overhead. Furthermore, parallel execution, knowledge distillation, and reinforcement learning will enhance agent efficiency, prevent latency bottlenecks and support long-term resilience.

\vspace{-3mm}
\subsection{Accuracy and performance evaluation}

We involved a performance evaluation considered 1) task completion rate, 2) factual consistency, and 3) semantic faithfulness. Expert scoring of outputs consistently exceeded 9/10 with nearly 100\% task completion, confirming high factual accuracy and logical consistency. Using GenAI in optical network automation faces hallucination problems, where the GenAI generates incorrect or fabricated information. In our implementation, this risk is mitigated by 1) focusing on instruction-generation and orchestration tasks with deterministic data, 2) retrieval-augmented generation from a structured Shared Pool, and 3) built-in validation agents and optional human expert checks for critical actions. GenAI is not utilized to replace precise numerical computation or existing coding solutions, instead focusing on facilitating task coordination and allocation, reducing human workload.

\vspace{-4mm}
\subsection{Multimodal applications}

The introduction of multimodal data processing is a natural extension of the framework's capabilities \cite{ref15}. Multimodal GenAI could simultaneously process network alarms/logs (text), field monitoring data (video), and performance metrics (value, graph, diagram). For example, it can analyze performance graphs, such as power profiles, optical time-domain reflectometer (OTDR) traces, eye diagrams, and constellation diagrams, enabling more comprehensive analysis and decision-making. Furthermore, the AI Agent with real-time data may include a 3-dimensional representation of the real-world in the near future, to ensure the health of the network without additional manual labor.

\vspace{-4mm}
\subsection{Security and explainability}

As GenAI systems take on increasingly critical roles in autonomous optical network operations, agents may access sensitive data, increasing privacy leakage risks. Our framework mitigates these risks through a hierarchical architecture with strict access control: only Network Director interacts with operators, AI Experts access network devices, and intermediate agents with shared resources operate internally with authentication protection. The Security Supporter verifies, analyzes, and authorizes configuration instructions before deployment. Future work can further enhance security and privacy by integrating privacy-preserving and encryption technologies, and AI models for detecting cyber-attacks or adversarial behaviors. Furthermore, explainability can be strengthened through interpretable outputs, such as transparent decision rationales, workflow visualizations, and security reports, enabling operators to monitor and build trust in zero-touch systems.

\vspace{-3mm}
\section{Conclusion}

This study explored the potential of GenAI to drive the autonomous operations for zero-touch optical networks using a hierarchical multi-agent framework. By incorporating specialized AI Agents, the framework was able to handle complex tasks, enhancing task allocation, coordination, implementation, evaluation, and summarization, thereby improving the overall efficiency of zero-touch management. We discussed the architecture and implementation of the framework, explaining how each component effectively coordination to support various operational tasks. Through three typical cases in a field-deployed network, covering the network planning, operation, and upgrade stages, we demonstrated the practical applications of the multi-agent framework in diverse optical network scenarios. Additionally, we discussed the challenges and opportunities associated with applying GenAI to the zero-touch optical network, with a focus on key issues such as accuracy, scalability, security, and adaptability. Our work provides an approach for implementing a GenAI-driven multi-agent system for optical network automation, which offers feasibility for future developments in the field.




\vspace{-12mm}

\begin{IEEEbiographynophoto}{Yao Zhang}
(zhang-yao@bupt.edu.cn) is a Ph.D. candidate at the Beijing University of Posts and Telecommunications (BUPT).
\end{IEEEbiographynophoto}

\vspace{-12.6mm}

\begin{IEEEbiographynophoto}{Yuchen Song}
(songyc@bupt.edu.cn) is a Ph.D. candidate at BUPT.
\end{IEEEbiographynophoto}

\vspace{-12.6mm}

\begin{IEEEbiographynophoto}{Shengnan Li}
(shengnanli@bupt.edu.cn) is a Ph.D. candidate at BUPT.
\end{IEEEbiographynophoto}

\vspace{-12.6mm}

\begin{IEEEbiographynophoto}{Yan Shi}
(shiyan49@chinaunicom.cn) currently works at China Unicom Research Institute. She received her Ph.D. degree from BUPT.
\end{IEEEbiographynophoto}

\vspace{-12.6mm}

\begin{IEEEbiographynophoto}{Shikui Shen}
(shensk@chinaunicom.cn) currently works at China Unicom Research Institute. He mainly focuses on the technique research and standardization of optical networks.
\end{IEEEbiographynophoto}

\vspace{-12.6mm}

\begin{IEEEbiographynophoto}{Xiongyan Tang}
(tangxy@chinaunicom.cn) currently works at China Unicom Research Institute. He is the Chief Scientist with China Unicom Research Institute and the Vice Dean of China Unicom Research Institute.
\end{IEEEbiographynophoto}

\vspace{-12.6mm}

\begin{IEEEbiographynophoto}{Min Zhang}
(mzhang@bupt.edu.cn) is currently a professor with BUPT. His main research interests include advanced optical communication systems and networks.
\end{IEEEbiographynophoto}

\vspace{-12.6mm}

\begin{IEEEbiographynophoto}{Danshi Wang}
(Senior Member, IEEE) (danshi\_wang@bupt.edu.cn) is currently a professor with the Institute of Information Photonics and Optical Communications, BUPT. He has published more than 170 articles. His research interests include AI for science, digital twin optical network.
\end{IEEEbiographynophoto}

\vfill

\end{document}